\documentclass[12pt]{article}
\usepackage{comment}
\usepackage{textcomp}
\usepackage{chetnew}
\usepackage{amssymb,amsmath,amsfonts,mathrsfs,braket}
\usepackage[utf8]{inputenc}
\usepackage{graphicx,longtable,afterpage}
\usepackage{cleveref}

\usepackage{cancel}

\usepackage[backend=biber,style=numeric-comp,sorting=none]{biblatex}
\usepackage{floatrow}
\newfloatcommand{capbtabbox}{table}[][\FBwidth]

\def\one{{\,\hbox{1\kern-.8mm l}}}

\newcommand{\CC}{\mathcal{C}}

\allowdisplaybreaks

\def\makeatletter{\catcode`\@=11}% 11:letter
\makeatletter
\def\mathbox#1{\hbox{$\m@th#1$}}%
\def\math@ccstyles#1#2#3#4#5#6#7{{\leavevmode
      \setbox0\mathbox{#6#7}%
      \setbox2\mathbox{#4#5}%
      \dimen@ #3%
      \baselineskip\z@\lineskiplimit#1\lineskip\z@
      \vbox{\ialign{##\crcr
             \hfil \kern #2\box2 \hfil\crcr
             \noalign{\kern\dimen@}%
             \hfil\box0\hfil\crcr}}}}
\def\mathaccstyles{\math@ccstyles\maxdimen}
\def\maththroughstyles{\math@ccstyles{-\maxdimen}}
\def\unity%
 {\maththroughstyles{.45\ht0}\z@\displaystyle {\mathchar"006C}\displaystyle 1}

\def\AA{{\cal A}}
\def\BB{{\cal B}}
\def\CC{{\cal C}}
\def\DD{{\cal D}}
\def\EE{{\cal E}}

\def\GG{{\cal G}}

\def\LL{{\cal L}}

\def\NN{{\cal N}}
\def\OO{{\cal O}}

\def\XX{{\cal X}}
\def\YY{{\cal Y}}

\def\beq{\begin{equation}}
\def\eeq{\end{equation}}
\newcommand{\bea}{\begin{eqnarray}}
\newcommand{\eea}{\end{eqnarray}}
\def\bal{\begin{align}}
\def\eal{\end{align}}

\preprint{\hfill QMUL-PH-26-14}

\title{\vspace{-1.cm} 
Reduced superblocks at next-to-next-to-extremality for half-maximally supersymmetric CFTs} 

\author{
 Mitchell~Woolley}

\affiliation{
 Centre for Theoretical Physics, Department of Physics and Astronomy\\ Queen Mary University of London, London E1 4NS, UK \vspace{0.3cm} $ $ \\
\vspace{0.3cm} $ $

\vspace{0.3cm}
{\tt \small
mitchell.woolley@qmul.ac.uk}}

\abstract{We consider mixed four-point correlators of 1/2-BPS operators $\phi_{k}$ in SCFTs with eight real Poincaré supercharges, namely the 3d $\mathcal{N}=4$, 4d $\mathcal{N}=2$, 5d $\mathcal{N}=1$, and 6d $\mathcal{N}=(1,0)$ theories. Using the basis of solutions to the superconformal Ward identity introduced in \cite{Dolan:2004mu}, we demonstrate that the dynamical data in mixed correlators of extremality $\mathcal{E}=2$ is encoded in certain simpler ``reduced correlator" functions that admit a block expansion, in close analogy to the recent result for maximally supersymmetric CFTs in \cite{Woolley:2026cii}. These reduced blocks similarly involve ordinary blocks with shifted kinematics and reproduce what is known in 4d, generalize a known example in 6d, and offer novel results in 3d and 5d.}

\date{\today}

\begin{document}

\maketitle

\hypersetup{pageanchor=true}

\setcounter{tocdepth}{2}

\toc 
\section{Introduction}
\label{intro}
The conformal bootstrap program aims to rigorously carve out the space of consistent conformal field theories (CFTs) by constraining their defining CFT data, namely scaling dimensions $\Delta_i$ and three-point coefficients $\lambda_{ijk}$ of operators $\OO_i$ using crossing symmetry and unitarity. Superconformal field theories (SCFTs) provide landmarks in this space of theories by their amenability to exact methods by virtue of supersymmetry. These theories can also provide organizing principles for this space, in that many families of SCFTs are putatively classified by their top-down realizations in string theory and M-theory. Such constructions also predict interacting SCFTs in $d=5,6$ whose existence, let alone their degrees of freedom, would have been difficult to deduce without this perspective.

While maximally supersymmetric CFTs, i.e. the 6d $\NN=(2,0)$, 4d $\NN=4$, and 3d $\NN=8$ SCFTs with sixteen real Poincaré supercharges provide a natural starting point in the superconformal bootstrap, the possibilities become far richer when considering theories with half of the maximal amount of supersymmetry, i.e. containing eight real Poincaré supercharges. In $d>2$, these are the\footnote{We note that in 5d, eight real Poincaré supercharges is in fact the maximum amount of supersymmetry to remain compatible with conformal symmetry.}
\begin{itemize}
    \item 6d $\NN=(1,0)$ SCFTs with superconformal algebra $\mathfrak{osp}(8^*|2)\supset\mathfrak{so}(6,2)\times\mathfrak{su}(2)_R$
    \item 5d $\NN=1$ SCFTs with superconformal algebra $\mathfrak{f}(4)\supset\mathfrak{so}(5,2)\times\mathfrak{su}(2)_R$
    \item 4d $\NN=2$ SCFTs with superconformal algebra $\mathfrak{su}(2,2|2)\supset\mathfrak{so}(4,2)\times\mathfrak{su}(2)_R\times\mathfrak{u}(1)_r$
    \item 3d $\NN=4$ SCFTs with superconformal algebra $\mathfrak{osp}(4|4)\supset\mathfrak{so}(3,2)\times\mathfrak{su}(2)_L\times\mathfrak{su}(2)_R$
\end{itemize}
In each of these SCFTs, the maximal bosonic subalgebra of the superconformal algebra contains an $\mathfrak{su}(2)_R$ R-symmetry factor which allows for a uniform treatment in this paper. Reduced supersymmetry opens the door to physically interesting features like the presence of global symmetries. In holographic examples, this paves the way to studying gluon scattering in gauge theories in AdS quantum gravity using the numerical and analytic bootstrap\footnote{See e.g. \cite{Chang:2017xmr,Chang:2017cdx,Beem:2014zpa,Chang:2019dzt} and \cite{Alday:2021odx,Chester:2023qwo,Behan:2024vwg,Chester:2025wti,Chester:2025jxg} for numerical and analytical bootstrap results in these directions, respectively.}.

A basic ingredient in the superconformal bootstrap is the decomposition of four-point correlators into superconformal blocks (superblocks) which package together the superconformal primary and its conformal descendants, together with superconformal descendant primaries and their conformal descendants. In the case of four-point correlators of 1/2-BPS operators, the determination of superblocks requires 1) the spectrum of conformal families in a given superconformal multiplet and 2) the proportionality constants that relate the three-point coefficients $\lambda_{ijk}$ of superconformal descendants to that of the superconformal primary. These calculations were initiated for such SCFTs in \cite{Dolan:2004mu} and pursued further in \cite{Bobev:2017jhk,Chang:2017xmr}, culminating in the computation of superblocks for all multiplets exchanged in four-point functions of 1/2-BPS operators of arbitrary $\mathfrak{su}(2)_R$ charge in \cite{Baume:2019aid}\footnote{We also point the reader to the alternative formulations of superconformal blocks presented in \cite{Buric:2019rms,Aprile:2021pwd}.}.

In this work, we will use the basis of solutions to the superconformal Ward identity introduced in \cite{Dolan:2004mu}, which was expressed in terms of an operator $\Delta_{\varepsilon}$ acting on a set of unconstrained two-variable functions $b^{\{k_i\}}_{n}$ and a one-variable function $f^{\{k_i\}}$, both of which we will refer to as ``reduced correlators". In direct analogy to the maximally supersymmetric study in \cite{Woolley:2026cii}, the novel outcome of this work will be to extract block decompositions of $b^{\{k_i\}}_{0}$ and $f^{\{k_i\}}$ for a class of next-to-next-to-extremal four-point functions $\langle\phi_{k_1}\phi_{k_2}\phi_{k_3}\phi_{k_1+k_2+k_3-4}\rangle$ in dimensions $d=2(\varepsilon+1)=3,4,5,6$\footnote{Note that there are additional subtleties in the 4d $\NN=2$ scenario stemming from the chiral algebra sub-sector discovered in \cite{Beem:2013sza}. For brevity, we refer the reader to \cite{Beem:2014zpa} and instead focus on aspects of superconformal blocks that hold in all $d=2(\varepsilon+1)=3,4,5,6$.}. These reduced blocks can be re-inserted into the $\mathfrak{su}(2)_R$ channel equations to generate the full superconformal blocks computed in \cite{Baume:2019aid}. Reduced blocks have previously been derived for the configuration $\langle\phi_{2}\phi_{2}\phi_{2}\phi_{2}\rangle$ in 4d in \cite{Dolan:2001tt,Beem:2014zpa} and in 6d in \cite{Chang:2017xmr}. Our results generalize these to account for more complicated configurations $\langle\phi_{k_1}\phi_{k_2}\phi_{k_3}\phi_{k_1+k_2+k_3-4}\rangle$ and extend them to odd dimensions $d=3,5$. 

We have arranged the remainder of this paper as follows. In Section \ref{setup}, we review a class of mixed four-point functions of 1/2-BPS operators and the superconformal Ward identity they obey. We describe the superconformal block decomposition that the Ward identity implies and generalize another method of satisfying the superconformal Ward identity put forth in \cite{Dolan:2004mu}. In Section \ref{E2} we relate these two formulations through a system of three $\mathfrak{su}(2)_R$ channel equations relating superblocks to combinations of the operator $\Delta_\varepsilon$ acting on the reduced correlators $b^{\{k_i\}}$ and $f^{\{k_i\}}$.  We demonstrate how these operator equations can be solved in terms of reduced block constituents of $b^{\{k_i\}}_0$ and $f^{\{k_i\}}$, which involve conformal blocks with shifted external kinematics. In Section \ref{outlook} we summarize our results and point toward future directions. Finally, we define our conformal block conventions in Appendix \ref{superblockconventions} and recall relevant properties of Jack polynomials in Appendix \ref{jackpolys}. Several additional technicalities concerning Jack polynomials and conformal block recursion relations appeared in the appendices of \cite{Woolley:2026cii} and we will often refer the reader to that work for brevity.

\section{Four-point functions of 1/2-BPS operators}
\label{setup}
In this section, we define the class of four-point functions we will consider, set conventions, and write down the crossing equations obeyed by these mixed correlators. We describe the superconformal Ward identity and two important decompositions of its solutions, namely an expansion in superblocks and the reduced correlator decomposition of \cite{Dolan:2004mu}. We also summarize the OPE superselection rules and resulting supermultiplets exchanged in our four-point functions, adopting the notation of \cite{Chang:2017xmr,Baume:2019aid}.

\subsection{Kinematics and crossing equations}
We will be concerned with four-point functions of 1/2-BPS superconformal primaries $\phi_k$ of the so-called $\DD[k]$ multiplets, which are isospin $\frac{k}{2}$ symmetric, traceless representations of the $\mathfrak{su}(2)_R$ factor of the R-symmetry group with protected conformal dimension $\Delta=\varepsilon k$\footnote{By labeling $\mathfrak{su}(2)_R$ irreps with $k=2J_R$ for half-integral $J_R$, we follow the conventions of \cite{Chang:2017xmr,Baume:2019aid} while drawing analogies with the maximally supersymmetric setup in \cite{Woolley:2026cii}.}\footnote{We point out that SCFTs with eight supercharges can admit additional R-symmetry factors (when $d<5$), as well as global symmetries. These will not play a role in this paper so we suppress their associated indices.}. We will eschew index clutter by contracting $\phi_{\alpha_1\dots \alpha_k}(x)$ with $SU(2)_R$ polarization spinors $y^\alpha$:
\begin{align}
    \phi_k(x,y)\equiv \phi_{\alpha_1\dots \alpha_k}(x)y^{\alpha_1}\cdots y^{\alpha_k}.
\end{align}
Superconformal symmetry then allows us to factorize four-point functions of 1/2-BPS operators into a kinematic factor and a dynamical function $\mathcal{G}^{k_1k_2k_3k_4}$ by writing
\begin{align}
    \langle \phi_{k_1}(x_1,y_1)\phi_{k_2}(x_2,y_2)\phi_{k_3}(x_3,y_3)\phi_{k_4}(x_4,y_4) \rangle =&\; \nonumber \\
    \left(\frac{y_{12}}{x^{2\varepsilon}_{12}}\right)^{\frac{k_1+k_2}{2}}
    \left(\frac{y_{34}}{x^{2\varepsilon}_{34}}\right)^{\frac{k_3+k_4}{2}}
    \left(\frac{y_{14}\,x_{24}^{2\varepsilon}}{y_{24}\,x_{14}^{2\varepsilon}}\right)^{\frac{k_{12}}{2}}&\;
    \left(\frac{y_{12}y_{34}}{y_{14}y_{24}}\right)^{\frac{k_{34}}{2}}
    \left(\frac{x_{14}^{2\varepsilon}x_{24}^{2\varepsilon}}{x_{12}^{2\varepsilon}x_{34}^{2\varepsilon}}\right)^{\frac{k_{34}}{2}}\mathcal{G}^{k_1k_2k_3k_4}(U,V;\alpha),
\label{eq:full4pt}
\end{align}
 where we consider operator orderings with $k_1,k_2,k_3\leq k_4$ and we define $x_{ij}=x_i-x_j$, $y_{ij}=\epsilon_{\alpha\beta}y_i^{\beta}y_j^{\alpha}$, and $k_{ij}=k_i-k_j$. The dynamical function $\mathcal{G}^{k_1k_2k_3k_4}$\footnote{To avoid clutter, we suppress the dependence of $\mathcal{G}^{k_1k_2k_3k_4}$ on $\varepsilon$, which will be made clear by the context.} depends on conformal cross-ratios $U$ and $V$ and the $\mathfrak{su}(2)_R$ cross-ratio $\alpha$ defined by\footnote{Recall that the $\mathfrak{su}(2)_R$ spinor identity $y_{14}y_{23}-y_{13}y_{24}+y_{12}y_{34}=0$ implies that we can only construct a single independent R-symmetry cross-ratio.}
 \begin{align}
    U=\frac{x_{12}^2x_{34}^2}{x_{13}^2x_{24}^2},\;\;\;\;V=\frac{x_{14}^2x_{23}^2}{x_{13}^2x_{24}^2},\;\;\;\;\alpha=\frac{y_{13}y_{24}}{y_{12}y_{34}}.
\end{align}
 This organization implies that $\mathcal{G}^{k_1k_2k_3k_4}$ is a polynomial in $\alpha$ of degree $\EE$, where the extremality $\EE$ is an integer obeying $k_i\geq\EE\geq0$ which quantifies the complexity of the $\mathfrak{su}(2)_R$ representations exchanged in $\GG^{\{k_i\}}$. Permuting the operators $\phi_{k_i}(x_i,y_i)$ in  full correlator \eqref{eq:full4pt} imposes crossing relations between different channels, namely
\begin{align}
\mathcal{G}^{k_1k_2k_3k_4}(U,V;\alpha)=&\;\mathcal{G}^{k_2k_1k_3k_4}\left(\frac{U}{V},\frac{1}{V};1-\alpha\right)
\label{eq:crossing12}\\
=&\;\left(\frac{U^{\varepsilon}}{V^\varepsilon}(1-\alpha)\right)^{\frac{k_1+k_2+k_3-k_4}{2}}\mathcal{G}^{k_3k_2k_1k_4}\left(V,U;\frac{\alpha}{\alpha-1}\right)
\label{eq:crossing13}\\
=&\;\left(U^\varepsilon\alpha\right)^{\frac{k_1+k_2+k_3-k_4}{2}}\mathcal{G}^{k_1k_3k_2k_4}\left(\frac{1}{U},\frac{V}{U};\frac{1}{\alpha}\right).
\label{eq:crossing23}
\end{align}
As usual, the condition derived by swapping $1\leftrightarrow2$ is manifestly satisfied term-by-term in the block expansions appearing below, while the remaining $1\leftrightarrow3$ and $2\leftrightarrow3$ crossing relations imply non-trivial constraints on four-point functions.

\subsection{Superconformal Ward identity, superblocks, and superconformal representation theory}
The bosonic subalgebra of the superconformal algebra allows us to decompose four-point correlators $\mathcal{G}^{k_1k_2k_3k_4}\equiv \mathcal{G}^{\{k_i\}}$ in a basis of exchanged irreducible representations (irreps) of the R-symmetry algebra $\mathfrak{su}(2)_R$. Taking $k_1\leq k_2$, the allowed $\mathfrak{su}(2)_R$ irrep exchanges in the OPE $\phi_{k_1}\times \phi_{k_2}$ are given by 
\begin{align}
    [k_1]\otimes [k_2]=&\;\bigoplus_{m=\frac{k_2-k_1}{2}}^{\frac{k_1+k_2}{2}} [2m].
    \label{eq:so3tp}
\end{align}
Focusing on the $s$-channel, the admissible $\mathfrak{su}(2)_R$ 
irreps exchanged in $\mathcal{G}^{\{k_i\}}$ are captured by $([k_1]\otimes [k_2])\cap([k_3]\otimes [k_4])$. To avoid unnecessarily many conditional expressions and without loss of generality, we will restrict to the class of configurations with $k_4=k_1+k_2+k_3-2\EE$. We do not impose any ordering among $k_1,k_2,k_3$ and these configurations are sufficient to consider all of the channels appearing in the crossing equations \eqref{eq:crossing12}-\eqref{eq:crossing23}. Temporarily specializing to $k_1\leq k_2$ again, we can decompose $\mathcal{G}^{\{k_i\}}$ into channels labeled by these irreps by writing 
\begin{align}
    \mathcal{G}^{\{k_i\}}(U,V;\alpha)=\sum_{m=\frac{k_1+k_2}{2}-\EE}^{\frac{k_1+k_2}{2}}\; P_{m-\frac{\kappa_t+\kappa_u}{4}}^{\left(\frac{\kappa_t}{2},\frac{\kappa_u}{2}\right)}\left(2\alpha-1\right)
    A_{2m}^{\{k_i\}}(U,V).
\label{eq:Adecomp}
\end{align}
where $P^{(a,b)}_{m}$ are Jacobi polynomials\footnote{Note that we normalize our Jacobi polynomials such that $P_m^{(a,b)}(1)=1$.} of degree $m$ in $\alpha$ that encode the exchange of $\mathfrak{su}(2)_R$ irreps with isospin $m$ and we define $\kappa_t=\left|k_{12}-k_{34}\right|$ and $\kappa_u=\left|k_{23}-k_{14}\right|$. We can then take the $s$-channel OPEs $\phi_{k_1}\times \phi_{k_2}$ and $\phi_{k_3}\times \phi_{k_4}$ and expand the R-symmetry channel functions $A_{2m}^{\{k_{i}\}}$ in the $2(\varepsilon+1)$-dimensional bosonic conformal blocks $G_{\Delta,\ell}^{\Delta_{12},\Delta_{34}}$ defined in Appendix \ref{superblockconventions} by writing
\begin{align}
    A_{2m}^{\{k_{i}\}}(U,V)=U^{\frac{\Delta_{34}}{2}}\sum_{\Delta,\ell}\lambda_{k_1k_2\mathcal{O}_{\Delta,\ell,2m}}\lambda_{k_3k_4\mathcal{O}_{\Delta,\ell,2m}}G_{\Delta,\ell}^{\Delta_{12},\Delta_{34}}(U,V).
\label{eq:Ablockdecomp}
\end{align}
The fermionic generators of each superconformal algebra in $d=2(\varepsilon+1)$ dimensions imposes additional constraints which are captured by the superconformal Ward identity \cite{Dolan:2004mu}
\begin{align}
    \left(z\partial_z-\varepsilon\alpha\partial_\alpha\right)\mathcal{G}^{\{k_i\}}\left(z,\bar z;\alpha\right)\bigg|_{\alpha\rightarrow z^{-1}}=0,
\label{eq:scwi}
\end{align}
and likewise for $z\rightarrow\bar{z}$, where $U=z \bar{z}$ and $V=(1-z)(1- \bar{z})$. This allows us to repackage $\mathcal{G}^{\{k_i\}}$ into an expansion in superblocks $\mathfrak{G}_\mathcal{X}^{\Delta_{12},\Delta_{34}}$, which individually satisfy \eqref{eq:scwi}. We can write  
\begin{align}
\mathcal{G}^{\{k_i\}}(U,V;\alpha)=\sum_{\mathcal{X}\in (\phi_{k_1}\times \phi_{k_2})\cap(\phi_{k_3}\times \phi_{k_4})}\lambda_{k_1 k_2 \mathcal{X}}\lambda_{k_3 k_4 \mathcal{X}}\;\mathfrak{G}_\mathcal{X}^{\Delta_{12},\Delta_{34}}(U,V;\alpha),
\label{eq:superblockdecomp}
\end{align}
where for $k_1\leq k_2$, superblocks take the form
\begin{align}
    \mathfrak{G}_\mathcal{X}^{\Delta_{12},\Delta_{34}}(U,V;\alpha)=&\;\sum_{m=\frac{k_1+k_2}{2}-\EE}^{\frac{k_1+k_2}{2}}\; P_{m-\frac{\kappa_t+\kappa_u}{4}}^{\left(\frac{\kappa_t}{2},\frac{\kappa_u}{2}\right)}\left(2\alpha-1\right)\sum_{\mathcal{O}\in\mathcal{X}} \mathcal{C}^{\mathcal{X},\Delta_{12},\Delta_{34}}_{\mathcal{O}_{\Delta,\ell,2m}}\;U^{\frac{\Delta_{34}}{2}}\;G_{\Delta,\ell}^{\Delta_{12},\Delta_{34}}(U,V)\nonumber \\
    \equiv&\;\sum_{m=\frac{k_1+k_2}{2}-\EE}^{\frac{k_1+k_2}{2}}\; P_{m-\frac{\kappa_t+\kappa_u}{4}}^{\left(\frac{\kappa_t}{2},\frac{\kappa_u}{2}\right)}\left(2\alpha-1\right)A_{2m;\Delta',\ell'}^{\{k_i\}}\left(U,V\right),
\label{eq:superblock}
\end{align}
where we defined $A_{2m;\Delta',\ell'}^{\{k_i\}}$ to be the channel contribution coming from the supermultiplet $\XX$, labelled by its (unique) primary $(\Delta',\ell')$ transforming in the highest $\mathfrak{su}(2)_R$ irrep. In this organization, supersymmetry has identified all products of three-point coefficients of superdescendant primaries $\mathcal{O}_{\Delta,\ell,2m}$ in a supermultiplet to be proportional to those of the superconformal primary, which we also denote $\mathcal{X}$. The proportionality constants $\mathcal{C}^{\mathcal{X},\Delta_{12},\Delta_{34}}_{\mathcal{O}_{\Delta,\ell,2m}}=\frac{\lambda_{k_1 k_2 \mathcal{O}_{\Delta,\ell,2m}}\lambda_{k_3 k_4 \mathcal{O}_{\Delta,\ell,2m}}}{\lambda_{k_1 k_2 \mathcal{X}}\lambda_{k_3 k_4 \mathcal{X}}}$ weight different primaries in a supermultiplet. All such coefficients (and therefore the structure of all superblocks) for any $\XX$ have been determined for four-point correlators in configurations with $k_1\leq k_2\leq k_3\leq k_4$ in \cite{Baume:2019aid}.

The superconformal multiplets $\mathcal{X}$ that can appear in the OPE $\phi_{k_1}\times \phi_{k_2}$ are dictated by the selection rules for 1/2-BPS operators in half-maximally supersymmetric CFTs determined e.g. in \cite{Ferrara:2001uj,Chang:2017xmr,Baume:2019aid}. We will follow the example of \cite{Chang:2017xmr,Baume:2019aid} and denote superconformal representations in 6d language, where superconformal multiplets are organized in the following families
\begin{align}
    \LL[2J_R]_{\Delta,\ell}:\hspace{1cm}\Delta>&\;2\varepsilon J_R+\ell+\mu\\
    \AA[2J_R]_{\Delta,\ell}:\hspace{1cm}\Delta=&\;2\varepsilon J_R+\ell+4\varepsilon-2\\
    \BB[2J_R]_{\Delta,\ell}:\hspace{1cm}\Delta=&\;2\varepsilon J_R+\ell+2\varepsilon\\
    \CC[2J_R]_{\Delta,\ell}:\hspace{1cm}\Delta=&\;2\varepsilon J_R+2\\
    \DD[2J_R]_{\Delta,\ell}:\hspace{1cm}\Delta=&\;2\varepsilon J_R
\end{align}
where $\varepsilon=\frac{d-2}{2}$ and the unprotected $\LL$-type multiplets are defined in terms of 
\begin{align}
    \mu=\begin{cases}
        4\varepsilon-2&\;\text{for}\hspace{1cm}4\leq d\leq6,
        \\
        2\varepsilon\hspace{1cm}&\;\text{for}\hspace{1cm}2< d\leq4.
    \end{cases}
\end{align}
\newpage
\noindent
Taking $k_1\leq k_2$, we can express the superselection rules in this language by writing \cite{Baume:2019aid}
\begin{align}
    \mathcal{D}[k_1]_{\varepsilon k_1,0}\otimes\mathcal{D}[k_2]_{\varepsilon k_2,0}=&\; \bigoplus_{m=\frac{k_2-k_1}{2}}^{\frac{k_1+k_2}{2}}\mathcal{D}[2m]_{2\varepsilon m,0} \nonumber \\
    \oplus&\;\bigoplus_{m=\frac{k_2-k_1}{2}}^{\frac{k_1+k_2-2}{2}}\bigoplus_{\ell=0}^{\infty}\mathcal{B}[2m]_{\varepsilon (2m+2)+\ell,\ell} 
    \label{eq:superselect}
    \\ \oplus&\;\bigoplus_{m=\frac{k_2-k_1}{2}}^{\frac{k_1+k_2-4}{2}}\bigoplus_{\ell=0}^{\infty}\bigoplus_{\Delta}\mathcal{L}[2m]_{\Delta>2\varepsilon m+\ell+\mu,\ell}. \nonumber
\end{align}
We see that $\AA$-type and $\CC$-type multiplets are incompatible with the superconformal Ward identity and do not contribute to four-point functions of $\DD$-type operators \cite{Ferrara:2001uj,Chang:2017xmr}. Moreover, the superselection rule \eqref{eq:superselect} only describes what is allowed by superconformal symmetry and additional principles may remove operators from this list. For instance, when $k_1=k_2$, Bose symmetry dictates that irreps $[2a]$ (with $k_1=a$ even) must have even/odd $\ell$ for even/odd $a$ (corresponding to a symmetric/antisymmetric $\mathfrak{su}(2)_R$ irrep).

\subsection{Dolan, Gallot, and Sokatchev's solution to the superconformal Ward identity}
\label{DGSdecomp}
The authors of \cite{Dolan:2004mu} proposed another formulation of solutions to the superconformal Ward identity \eqref{eq:scwi} for four-point correlators $\langle\phi_k\phi_k\phi_k\phi_k\rangle$ in all SCFTs with R-symmetry algebra $\mathfrak{su}(2)_R$. They demonstrate that the superconformal Ward identity is satisfied by a sum of two-variable functions $b^{\{k_i\}}_{I}$ and a single-variable function $f^{\{k_i\}}$, acted on by differential operators. Apart from the extremality $\EE$, the superconformal Ward identities are insensitive to the kinematical configurations of the four-point correlators $\GG^{\{k_i\}}$ they constrain, allowing us to generalize their result by writing
\begin{align}
    \GG^{\{k_i\}}\left(U,V;\alpha\right)=&\;
    \sum_{I=0}^{\EE-2}U^{\varepsilon(I+2)}\alpha^I\Delta_{\varepsilon}\left[\left(z\alpha-1\right)\left(\bar{z}\alpha-1\right)b^{\{k_i\}}_{I}\left(U,V\right)\right]\nonumber\\
    +&\;U^\varepsilon\left(D_\varepsilon\right)^{\varepsilon-1}\left[\frac{(z \alpha-1)f^{\{k_{i}\}}(z)-(\bar{z} \alpha-1)f^{\{k_{i}\}}(\bar{z})}{z-\bar{z}}\right],
    \label{eq:DGSdecomphalfsusy}
\end{align}
in terms of the operators
\begin{align}
    \Delta_{f}=\left(D_{\varepsilon}\right)^{f-1}U^{f-1},\hspace{2cm}
    D_{\varepsilon}=\partial_ z\partial_{\bar{z}}-\frac{\varepsilon}{z-\bar{z}}\left(\partial_{z}-\partial_{\bar{z}}\right).
\label{eq:Depsilon}
\end{align}
Other than when $\varepsilon=1,2,3,\dots$ (corresponding to even number of dimensions $d\geq4$), the fractional and/or negative powers of $D_\varepsilon$ render the operator $\Delta_\varepsilon$ non-local. To manipulate these expressions, it is convenient to note that $\Delta_\varepsilon$ has as its eigenfunctions a family of two-variable symmetric functions $P^{(\varepsilon)}_{a,b}$ known as Jack polynomials \cite{Jack:1970,Dolan:2000ut,Dolan:2011dv} which satisfy
\begin{align}
    \Delta_f P^{(\varepsilon)}_{a,b}(z,\bar{z})=(a+\varepsilon+1)_{f-1}(b+1)_{f-1}P^{(\varepsilon)}_{a,b}(z,\bar{z}),
\end{align}
in terms of the Pochhammer symbol $(x)_y=\Gamma[x+y]/\Gamma[x]$. The operator $\Delta_\varepsilon$ has a non-trivial kernel and great care needs to be taken to unambiguously constrain the reduced correlator functions, e.g. through the formulation of reduced crossing equations. It will also be useful to note that $2(\varepsilon+1)$-dimensional conformal blocks $G_{\Delta,\ell}^{\Delta_{12},\Delta_{34}}$ can also be expanded in Jack polynomials as
\begin{align}
    G_{\Delta,\ell}^{\Delta_{12},\Delta_{34}}(z,\bar{z})=(-1)^\ell\sum_{m=0}^\infty\sum_{n=0}^\infty r_{m,n;\Delta,\ell}^{\Delta_{12},\Delta_{34}}\;P^{(\varepsilon)}_{\frac{\Delta+\ell}{2}+m,\frac{\Delta-\ell}{2}+n}(z,\bar{z}).
\end{align}
In Appendix \ref{jackpolys}, we collect several properties of Jack polynomials and the expansion coefficients $r_{m,n;\Delta,\ell}^{\Delta_{12},\Delta_{34}}$ that will be essential for the results of this paper. 

\section{Reduced blocks at next-to-next-to-extremality}
\label{E2}
In this section, we specialize to four-point correlators $\langle\phi_{k_1}\phi_{k_2}\phi_{k_3}\phi_{k_1+k_2+k_3-4}\rangle$ with extremality $\EE=2$. We will outline the representation theory of these correlators and relate the superblock and reduced correlator decompositions in Equations \eqref{eq:superblockdecomp} and \eqref{eq:DGSdecomphalfsusy} via R-symmetry channel equations. We solve these equations by presenting block expansions of the reduced correlator functions $b^{\{k_i\}}$ and $f^{\{k_i\}}$ in terms of ordinary $2\left(\varepsilon+1\right)$-dimensional bosonic blocks and global $SL(2,\mathbb{R})$ blocks for each $\varepsilon$. Finally, we associate $b^{\{k_i\}}_{\Delta,\ell}$ and the single-variable reduced blocks $f^{\{k_i\}}_{\ell}$ with the multiplets appearing in $\langle \phi_{k_1}\phi_{k_2}\phi_{k_3}\phi_{k_1+k_2+k_3-4}\rangle$.

\subsection{$\frak{su}(2)_R$ channel equations}
\label{E2Rchannels}

The tensor products in \eqref{eq:so3tp} demonstrate that correlators in this configuration will exchange three $\mathfrak{su}(2)_R$ irreps. We define the combinations
\begin{align}
    m\left(\EE\right)=\frac{1}{2}\min\left(k_1+k_2,k_3+k_4\right), \hspace{1cm}
p\left(\EE\right)=\min\left(k_1+k_2,k_3+k_4\right)-2\EE,
    \label{eq:mpE}
\end{align}
which will repeatedly appear in various quantum numbers. For our $\EE=2$ configuration, we will abbreviate $p(\EE=2)=k_1+k_2-4\equiv p$, and the exchanged irreps are listed as
\begin{align}
    ([k_1]\otimes [k_2])\cap([k_3]\otimes [k_4])=
    &\;[p]\oplus[p+2]\oplus[p+4],
    \label{eq:sodtpE2}
\end{align}
where the irreps $[p]$ and $[p+4]$ have the opposite parity to $[p+2]$. This distinction will be important for configurations where one or more of the OPEs involve operators with identical dimension, in which case Bose symmetry restricts the parity of exchanged Lorentz irreps. In the configuration $\langle\phi_{k_1}\phi_{k_2}\phi_{k_3}\phi_{k_1+k_2+k_3-4}\rangle$, this only comes into effect when $k_1=k_2=2$. 

When $\EE=2$, the decomposition in Equation \eqref{eq:DGSdecomphalfsusy} involves one two-variable reduced correlator $b^{\{k_i\}}_{0}\equiv b^{\{k_i\}}$ and one single-variable reduced correlator $f^{\{k_i\}}$:
\begin{align}
    \GG^{\{k_i\}}\left(U,V;\alpha\right)=&\;
    U^{2\varepsilon}\Delta_{\varepsilon}\left[\left(z\alpha-1\right)\left(\bar{z}\alpha-1\right)b^{\{k_i\}}\left(U,V\right)\right]\nonumber\\
    +&\;U^\varepsilon\left(D_\varepsilon\right)^{\varepsilon-1}\left[\frac{(z \alpha-1)f^{\{k_i\}}(z)-(\bar{z} \alpha-1)f^{\{k_i\}}(\bar{z})}{z-\bar{z}}\right],
    \label{eq:DGSdecompE2}
\end{align}
We determine how $b^{\{k_i\}}$ and $f^{\{k_i\}}$ contribute to each channel by comparing Equations \eqref{eq:Adecomp} and \eqref{eq:DGSdecompE2}. In terms of $m(\EE)=\frac{k_1+k_2}{2}\equiv m$ and suppressing dependency on $U,V$, we find 
\begin{align}
    A^{\{k_{i}\}}_{2m}=&\;\frac{1}{\mathcal{Y}_{2m}^{\{k_i\}}}U^{2\varepsilon}\Delta_\varepsilon\left[Ub^{\{k_i\}}\right], 
    \label{eq:A1}\\
    A^{\{k_{i}\}}_{2m-2}=&\;\frac{1}{\mathcal{Y}_{2m-2}^{\{k_i\}}}U^{2\varepsilon}\Delta_\varepsilon\left[\left(
    \frac{\Delta _{12}}{\Delta _{34}-4 \varepsilon }
    +\frac{V-1}{U}\right)Ub^{\{k_i\}}\right]+\frac{1}{\mathcal{Y}_{2m-2}^{\{k_i\}}}U^{\varepsilon}\left(D_\varepsilon\right)^{\varepsilon-1}\left[\frac{zf^{\{k_i\}}(z)-\bar{z}f^{\{k_i\}}(\bar{z})}{z-\bar{z}}\right], 
    \label{eq:A2}\\
    A^{\{k_{i}\}}_{2m-4}=&\;U^{2\varepsilon}\Delta_\varepsilon\left[\left(\frac{\left(\Delta _{12}-\Delta _{34}+2\varepsilon\right) \left(\Delta _{12}+\Delta _{34}-2\varepsilon\right)}{4 \left(\Delta _{34}-2 \varepsilon\right) \left(\Delta_{34}-3 \varepsilon\right)}
    +\frac{\Delta _{12}}{2 \left(\Delta _{34}-2 \varepsilon\right)}\frac{V-1}{U}+\frac{V+1}{2U}
    \right)Ub^{\{k_i\}}\right]\nonumber\\
    +&\;U^{\varepsilon}\left(D_\varepsilon\right)^{\varepsilon-1}\Bigg[\frac{\Delta _{12}}{2 \left(\Delta _{34}-2 \varepsilon \right)}\frac{zf^{\{k_i\}}(z)-\bar{z}f^{\{k_i\}}(\bar{z})}{z-\bar{z}}
    +\frac{(z-2)f^{\{k_i\}}(z)-(\bar{z}-2)f^{\{k_i\}}(\bar{z})}{2(z-\bar{z})}\Bigg],
    \label{eq:A3}
    \end{align}
where we define
\begin{align}
    \mathcal{Y}_{q}^{\{k_i\}}=\frac{q!\left(\frac{k_{12}-k_{34}}{2}\right)!}{\left(\frac{q+k_{12}}{2}\right)!\left(\frac{q-k_{34}}{2}\right)!},
\label{eq:Rgluing}
\end{align}
\begin{comment}
where we define 
 and following \cite{Alday:2021odx}, we define a kinematical factor 
\begin{align}
    \YY_{m}^{\{k_i\}}=\frac{1}{m!}\Gamma\left[\frac{2m-\kappa_t-\kappa_u+4}{4}\right]\Gamma\left[\frac{2m+\kappa_t+\kappa_u+4}{4}\right],
\end{align}
in terms of $\kappa_t=\left|k_{12}-k_{34}\right|$ and $\kappa_u=\left|J_{21}-k_{34}\right|$.
\end{comment}
where we note that $\mathcal{Y}_{2m-4}^{\{k_i\}}=1$. Upon expanding $\GG^{\{k_i\}}$ in the reduced correlator basis proposed by \cite{Dolan:2004mu} and comparing with an expansion in $\mathfrak{su}(2)_R$ harmonic polynomials $P_m^{(a,b)}$, these channel equations have the same functional dependence on the two-variable reduced correlator $b^{\{k_i\}}$ as the top three channel equations appearing in the maximally supersymmetric case in \cite{Woolley:2026cii}. We have organized each $A^{\{k_{i}\}}_{2m}$ into sums of the operator combinations 
\begin{align}
    \left\{\Delta_\varepsilon,\Delta_\varepsilon\frac{V-1}{U},\Delta_\varepsilon\frac{V+1}{2U}\right\}\hspace{0.5cm}\text{and}\hspace{0.5cm}
    \left\{\left(D_\varepsilon\right)^{\varepsilon-1}\frac{z}{z-\bar{z}},\left(D_\varepsilon\right)^{\varepsilon-1}\frac{z-2}{2(z-\bar{z})}\right\}
\label{eq:recuringredients}
\end{align}
acting on $Ub^{\{k_{i}\}}$ and $f^{\{k_i\}}$ in anticipation of the fact that for appropriate ``reduced block" functions $b^{\{k_{i}\}}_{\XX}$ and $f^{\{k_i\}}_{\XX}$, each expression generates the correct linear combination of $2(\varepsilon+1)$-dimensional conformal blocks to realize the superconformal blocks computed in \cite{Baume:2019aid}.

\subsection{Reduced blocks}
\label{reducedblocks}
Given the channel equations \eqref{eq:A1}-\eqref{eq:A3}, we wish to solve them for ``reduced block" contributions to $b^{\{k_i\}}$ and $f^{\{k_i\}}$ that satisfy this system of equations for each supermultiplet $\XX$. First, we note that each $\XX$ contains a unique conformal primary transforming in the supermultiplet's highest $\mathfrak{su}(2)_R$ irrep. The resulting R-symmetry channel equation relates this conformal block to a reduced block contribution from $b^{\{k_i\}}$ or $f^{\{k_i\}}$ by the action of differential operators. We avoid difficulties in interpreting the operator $\Delta_\varepsilon$ by expanding the conformal block in Jack polynomials, which are eigenfunctions of $\Delta_\varepsilon$. To this end, we decompose the channel functions into sums of supermultiplet contributions by writing
\begin{align}
    A_{2m}^{\{k_{i}\}}\left(U,V\right)=\sum_{\mathcal{X}\in (\phi_{k_1}\times \phi_{k_2})\cap(\phi_{k_3}\times \phi_{k_4})}\lambda_{k_1 k_2 \mathcal{X}}\lambda_{k_3 k_4 \mathcal{X}}\;A_{2m;\Delta_\XX',\ell_\XX'}^{\{k_{i}\}}\left(U,V\right).
\end{align}
To interpret the sum over $\XX$, we use the superselection rule in \eqref{eq:superselect} on the OPEs $\phi_{k_1}\times\phi_{k_2}$ and $\phi_{k_3}\times\phi_{k_1+k_2+k_3-4}$ to determine that correlators $\langle \phi_{k_1}\phi_{k_2}\phi_{k_3}\phi_{k_1+k_2+k_3-4}\rangle$ will exchange the supermultiplets 
\begin{align}
    \mathcal{D}[p]_{\varepsilon p,0},\;\mathcal{D}[p+2]_{\varepsilon(p+2),0},\;\mathcal{D}[p+4]_{\varepsilon(p+4),0},\;\mathcal{B}[p]_{\varepsilon (p+2)+\ell,\ell},\;\mathcal{B}[p+2]_{\varepsilon (p+4)+\ell,\ell},\;\mathcal{L}[p]_{\Delta,\ell}, 
\label{eq:E2superselect}
\end{align}
again in terms of $p=k_1+k_2-4$. We further decompose these channel functions into reduced correlator contributions
\begin{align}
    A_{2m;\Delta_\XX',\ell_\XX'}^{\{k_{i}\}}\left(U,V\right)=A_{2m;\Delta_\XX',\ell_\XX'}^{\{k_{i}\};b}\left(U,V\right)+A_{2m;\Delta_\XX',\ell_\XX'}^{\{k_{i}\};f}\left(U,V\right).
\label{eq:Amsplit}
\end{align}
The strategy is now to identify the appearance of reduced correlator functions in the highest $\mathfrak{su}(2)_R$ channel with a unique conformal primary $\left(\Delta_\XX',\ell_\XX'\right)$\footnote{The primes in these quantum numbers serve to remind the reader that aside from $\DD$-type multiplets, these conformal primaries will not coincide with the superconformal primary.}, which depends on a given supermultiplet $\XX$. Given \eqref{eq:E2superselect}, we consider the expressions
\begin{align}
    A^{\{k_{i}\};b}_{2m;\Delta_\XX',\ell_\XX'}=&\;\frac{1}{\mathcal{Y}_{2m}^{\{k_i\}}}U^{2\varepsilon}\Delta_\varepsilon\left[Ub^{\{k_{i}\}}_\XX\right]=\mathcal{C}^{\mathcal{X},\Delta_{12},\Delta_{34}}_{\mathcal{O}_{\Delta_\XX',\ell_\XX',2m}}\;U^{\frac{\Delta_{34}}{2}}\;G^{\Delta_{12},\Delta_{34}}_{\Delta_\XX',\ell_\XX'} 
    \label{eq:bopeq}\\ &\;\hspace{1cm}\text{for} \hspace{1cm} \XX\in\left\{\mathcal{D}[p+4]_{\varepsilon(p+4),0},\;\mathcal{B}[p+2]_{\varepsilon (p+4)+\ell,\ell},\;\mathcal{L}[p]_{\Delta,\ell} \right\}, \nonumber \\
    \text{and separately}\hspace{1cm}&\;\nonumber \\
    A^{\{k_{i}\};f}_{2m-2;\Delta_\XX',\ell_\XX'}=&\;\frac{1}{\mathcal{Y}_{2m-2}^{\{k_i\}}}U^{\varepsilon}\left(D_\varepsilon\right)^{\varepsilon-1}\left[\frac{zf_\XX^{\{k_i\}}(z)-\bar{z}f_\XX^{\{k_i\}}(\bar{z})}{z-\bar{z}}\right]=\mathcal{C}^{\mathcal{X},\Delta_{12},\Delta_{34}}_{\mathcal{O}_{\Delta_\XX',\ell_\XX',2m-2}}\;U^{\frac{\Delta_{34}}{2}}\;G^{\Delta_{12},\Delta_{34}}_{\Delta_\XX',\ell_\XX'}
    \nonumber \\ &\;\hspace{1cm}\text{for} \hspace{1cm} \XX\in\left\{\mathcal{D}[p+2]_{\varepsilon(p+2),0},\;\mathcal{B}[p]_{\varepsilon (p+2)+\ell,\ell}\right\}.
\end{align}
Using similar manipulations to those in \cite{Woolley:2026cii}, we invert these expressions to deduce that the functions $b^{\{k_i\}}$ and $f^{\{k_i\}}$ admit expansions in the reduced blocks
\begin{align}
    b^{\{k_i\}}_{\Delta,\ell}(U,V)=&\;\frac{\Gamma \left[\frac{\Delta+\ell+\Delta_{34}-2(\varepsilon-1)}{2}\right]\Gamma \left[\frac{\Delta-\ell+\Delta_{34}-2(2\varepsilon-1)}{2}\right]}{\Gamma\left[\frac{\Delta+\ell+\Delta_{34}}{2}\right] \Gamma \left[\frac{\Delta -\ell+\Delta _{34}-2 \varepsilon }{2}\right]}\;U^{\frac{\Delta_{34}}{2}-2\varepsilon-1}\;G_{\Delta,\ell}^{\Delta_{12},\Delta_{34}-2(\varepsilon-1)}\left(U,V\right),
    \label{eq:breducedblock}\\
    f^{\{k_i\}}_{\ell}(z)=&\;(-1)^{\varepsilon+1}\frac{\Gamma \left[\ell+1\right]}{\Gamma [\varepsilon] \Gamma\left[\ell+\varepsilon\right]}z^{\frac{\Delta _{34}}{2}+\varepsilon-2}g^{\Delta_{12},\Delta_{34}-2(\varepsilon-1)}_{2\varepsilon-\Delta_{34}+\ell,\ell}(z),
\end{align}
where $G^{\Delta_{12},\Delta_{34}}_{\Delta,\ell}$ are $2(\varepsilon+1)$-dimensional bosonic blocks and $g^{\Delta_{12},\Delta_{34}}_{\Delta,\ell}$ denote the global $SL(2,\mathbb{R})$ blocks defined in Equation \eqref{eq:sl2Rblock}\footnote{The factors of $(-1)^{\varepsilon}$ appearing in $f^{\{k_i\}}_{\ell}$ follow from Jack polynomial identities in \cite{Dolan:2004mu}. They are part of the formal definition of the reduced blocks and do not violate unitarity, e.g. by affecting the reality of the OPE coefficients they multiply.}. With these ingredients in place, one expresses each reduced correlator as the sum
\begin{align}
    b^{\{k_i\}}(U,V)=&\;\sum_{\mathcal{X}\in (\phi_{k_1}\times \phi_{k_2})\cap(\phi_{k_3}\times \phi_{k_4})}\lambda_{k_1 k_2 \mathcal{X}}\lambda_{k_3 k_4 \mathcal{X}}\;\NN_{\XX}\;b^{\{k_i\}}_{\Delta_\XX,\ell_\XX}\left(U,V\right), \\
f^{\{k_{i}\}}(z)=&\;\sum_{\mathcal{X}\in (\phi_{k_1}\times \phi_{k_2})\cap(\phi_{k_3}\times \phi_{k_4})}\lambda_{k_1 k_2 \mathcal{X}}\lambda_{k_3 k_4 \mathcal{X}}\;\NN_\XX\;f^{\{k_{i}\}}_{\ell_\XX}\left(z\right).
\end{align}
The superconformal Ward identity only determines superblocks up to an overall normalization, which we denote by $\NN_\XX$. We will choose $\NN_\XX$ such that the superconformal primary block has unit coefficient. All that remains is to associate reduced blocks with supermultiplets $\XX$ appearing in \eqref{eq:E2superselect}, as we summarize in Table \ref{reducedblocktable}. The recursion relations in the appendices of \cite{Woolley:2026cii} demonstrate that upon insertion in the channel equations \eqref{eq:A1}-\eqref{eq:A3}, these reduced blocks generate the full superblocks computed in \cite{Baume:2019aid}\footnote{The superblocks and the normalization coefficient $\NN_\XX$ are convention-dependent. We use the same conventions as in \cite{Woolley:2026cii}.} so that in principle, no constraints are lost by bootstrapping the functions $b^{\{k_i\}}$ and $f^{\{k_i\}}$. 

\begin{table}[h]
\begin{tabular}{|l||l|l|l|l|}
\hline Supermultiplet $\XX$ & $b^{\{k_i\}}_{\Delta,\ell}(U,V)$ & $f^{\{k_i\}}_{\ell}(z)$ & $\NN_\XX$\\
\hline \hline 
$\mathcal{L}[p]_{\Delta,\ell}$ & $b^{\{k_i\}}_{\Delta+2,\ell}$ & --- & $\frac{\left(\Delta+\ell +\Delta _{34}\right) \left(\Delta-\ell+\Delta _{34}-2 \varepsilon \right)}{\left(\Delta+\ell +\Delta _{34}-2 \varepsilon+2\right)\left(\Delta-\ell+\Delta _{34}-4 \varepsilon+2\right)}$ \\
\hline $\mathcal{B}[p+2]_{\varepsilon (p+4)+\ell,\ell}$ & $b^{\{k_i\}}_{\varepsilon (p+4)+\ell+1,\ell+1}$ & --- & $\frac{2 \left(2 \varepsilon -\Delta _{34}\right)}{\left(\Delta _{12}-\Delta _{34}+2 \varepsilon \right)}\frac{\ell+\varepsilon}{\ell+1}$ \\
\hline $\mathcal{B}[p]_{\varepsilon (p+2)+\ell,\ell}$ & --- & $f^{\{k_i\}}_{\ell+1}$ & $\frac{\ell+\varepsilon}{\ell+1}$ \\
\hline $\mathcal{D}[p+4]_{\varepsilon(p+4),0}$ & $b^{\{k_i\}}_{\varepsilon (p+4),0}$ & --- & $\frac{4 \left(3 \varepsilon -\Delta _{34}\right) \left(4 \varepsilon -\Delta _{34}\right)}{\left(\Delta _{12}-\Delta _{34}+2 \varepsilon \right) \left(\Delta _{12}-\Delta _{34}+4 \varepsilon \right)}$\\
\hline $\mathcal{D}[p+2]_{\varepsilon(p+2),0}$ & --- & $f^{\{k_i\}}_{0}$ & $\frac{2 \left(2 \varepsilon -\Delta _{34}\right)}{\Delta _{12}-\Delta _{34}+2 \varepsilon }$ \\
\hline $\mathcal{D}[p]_{\varepsilon p,0}$ & --- & $(-1)^{\varepsilon+1}z^{-1}$ & 1 \\
\hline
\end{tabular}
\caption{The organization of normalized reduced blocks in extremality $\EE=2$ correlators.}
\label{reducedblocktable}
\end{table}

The non-trivial kernel of $\Delta_\varepsilon$ necessitates the consideration of uniqueness of the reduced block $b^{\{k_i\}}_{\Delta,\ell}$, which was derived as the inhomogeneous solution to the operator equation \eqref{eq:bopeq}. Any additional reduced block $\tilde{b}^{\{k_i\}}_{\Delta,\ell}$ would be a homogeneous solution satisfying
\begin{align}
    \Delta_\varepsilon\left[U \tilde{b}^{\{k_i\}}_{\Delta,\ell}\left(U,V\right)\right]=0.
\end{align}
Solutions to this equation involve the restricted Jack polynomial expansions that are used to define $f^{\{k_i\}}_{\ell}$ and inserting these into the other channels generates unphysical twist $t=\Delta-\ell=2\varepsilon-\Delta_{34}$ and $t=-\Delta_{34}$ blocks, leading us to conclude that $\tilde{b}^{\{k_i\}}_{\Delta,\ell}=0$ for physical four-point correlators. We point the reader to \cite{Woolley:2026cii} for a more detailed argument of the identical consideration in the maximally supersymmetric context.

The organization of reduced blocks in Table \ref{reducedblocktable} assigns the single-variable blocks to supermultiplets containing twist $t=\Delta-\ell=2\varepsilon-\Delta_{34}$ and $t=-\Delta_{34}$ operators. The derivation in \cite{Dolan:2004mu} shows that the term involving $f^{\{k_i\}}$ in \eqref{eq:DGSdecomphalfsusy} resembles the solution to the Ward identity for next-to-extremal correlators and indeed, $\EE=1$ correlators are comprised of operators with twists $t=2\varepsilon-\Delta_{34},-\Delta_{34}$. In $d=4$ $\NN=2$ SCFTs, this single-variable function is intimately tied to the chiral algebra subsector discovered in \cite{Beem:2013sza}; however, we emphasize that even without such a subsector in SCFTs with $\varepsilon\neq1$, the single-variable function plays a crucial role. We also point out that unlike in the maximally supersymmetric cases with $\varepsilon=1,2$, the single-variable functions are not manifestly determinable using crossing symmetry alone (unless $\varepsilon=1$). The existence of just a single R-symmetry cross-ratio restricts us from devising twists that isolate holomorphic constraints on $f^{\{k_i\}}$ and moreover, the operator $\Delta_\varepsilon$ makes the formulation of reduced crossing relations delicate, as we will discuss in Section \ref{outlook}.

\begin{comment}
\subsection{Reduced crossing equations}

\begin{align}
    0=&\;U^\varepsilon\Delta _{\varepsilon}\left[b(U,V)-\left(\frac{U}{V}\right)^{1-\varepsilon} b(V,U)\right]\nonumber\\-&\;\Delta _{\varepsilon}\left[\frac{U^{1-\varepsilon } (\mathit{f}(z)-\mathit{f}(\bar{z}))}{z-\bar{z}}\right]-\left(\frac{U}{V}\right)^{\varepsilon}\Delta _{\varepsilon }\left[\frac{U^{1-\varepsilon} (\mathit{f}(1-z)-\mathit{f}(1-\bar{z}))}{z-\bar{z}}\right],\\
    0=&\;\Delta _{\varepsilon}\left[U \left(b(U,V)-\left(\frac{U}{V}\right)^{1-\varepsilon } b(V,U)\right)\right]-V^{-\varepsilon } \Delta _{\varepsilon }\left[\frac{U^{1-\varepsilon } (z \mathit{f}(1-z)-\bar{z} \mathit{f}(1-\bar{z}))}{z-\bar{z}}\right],
    \\
    0=&\;U^\varepsilon\Delta _{\varepsilon }\left[V \left(b(U,V)-\left(\frac{U}{V}\right)^{1-\varepsilon } b(V,U)\right)\right]-\Delta _{\varepsilon }\left[\frac{U^{1-\varepsilon  } ((z-1) \mathit{f}(z)-(\bar{z}-1) \mathit{f}(\bar{z}))}{z-\bar{z}}\right].
\end{align}
\end{comment}

\section{Conclusions and Outlook}
\label{outlook}
The result of this paper was to compute reduced block decompositions for four-point functions $\langle\phi_{k_1}\phi_{k_2}\phi_{k_3}\phi_{k_1+k_2+k_3-4}\rangle$ of 1/2-BPS operators $\phi_{k_i}$ in the half-maximally supersymmetric CFTs in $d=3,4,5,6$ in terms of a single $2(\varepsilon+1)$-dimensional conformal block and a global $SL(2,\mathbb{R})$ block. The availability of reduced blocks should, in principle, present a drastic simplification of numerical and analytic superconformal bootstrap studies by allowing one to examine correlators in terms of substantially simpler block expansions. In the following, we describe questions concerning implementation and comment on further directions.

An important open problem involves the imposition of the crossing equations \eqref{eq:crossing12}-\eqref{eq:crossing23} as constraints on the reduced correlators themselves. The existence of just a single R-symmetry cross-ratio $\alpha$ in the half-maximally supersymmetric context renders it impossible to algebraically eliminate the action of $\Delta_\varepsilon$ even when it is a well-defined differential operator in integer $\varepsilon>0$. While this is no issue in 4d where $\Delta_1=1$, care needed to be taken in the 6d study of $\langle\phi_2\phi_2\phi_2\phi_2\rangle$ in \cite{Chang:2017xmr}, where the most general reduced crossing equation turned out to be inhomogeneous, i.e. it included an additional $z,\bar{z}$-dependent term\footnote{This term was forced to vanish in the case of 6d generalized free field theory.}. The situation is more severe in odd dimensions, where $\Delta_{\varepsilon}$ is only formally defined by its action on the functional space spanned by Jack polynomials. Not only does one need to account for the non-trivial kernel, one must also relate $\Delta_{\varepsilon}$ to its crossed image. For the crossing relations \eqref{eq:crossing12} and \eqref{eq:crossing23}, it is helpful to note that
\begin{align}
    \Delta_\varepsilon\Bigg|_{\substack{U\rightarrow\frac{U}{V}\\V\rightarrow\frac{1}{V}}}=V^{2\varepsilon}\Delta_\varepsilon V^{-\varepsilon-1},\hspace{1.5cm}\Delta_\varepsilon\Bigg|_{\substack{U\rightarrow V\\V\rightarrow U}}=\Delta_\varepsilon \left(\frac{U}{V}\right)^{1-\varepsilon},\hspace{1.5cm}\Delta_\varepsilon\Bigg|_{\substack{U\rightarrow\frac{1}{U}\\V\rightarrow\frac{V}{U}}}=U^{2\varepsilon}\Delta_\varepsilon U^{-2\varepsilon}.
\end{align}   

Another question arises in the case of $\varepsilon=\frac{1}{2}$, where the authors of \cite{Chester:2014mea} demonstrated that upon taking a certain twist, restricting operator insertions to a line devolves 3d $\NN=4$ correlators to those in a theory of topological quantum mechanics. The only contributions that survive this procedure are $\DD$-type operators, some of which are encoded by $f^{\{k_i\}}$ in Table \ref{reducedblocktable}. In analogy to the maximally supersymmetric setup in \cite{Woolley:2026cii}, the decomposition in \eqref{eq:DGSdecomphalfsusy} does not clearly delineate contributions that play a role in this twisted limit. It would be interesting to understand the behavior of the operator $\Delta_{\frac{1}{2}}$ in this limit or otherwise, devise an alternative basis of reduced correlators that makes the outcome of this twist manifest. A resolution to this issue, as well as an understanding of the crossing properties of $\Delta_\frac{1}{2}$, should enable one to reproduce the 1d sum rules for OPE coefficients that appeared in \cite{Chang:2019dzt}, but now using reduced correlators.

The reduced blocks presented in this work can be compared with the reduced correlators derived in Mellin space in \cite{Chester:2025jxg} for half-maximally supersymmetric CFTs in $3\leq d\leq 6$ using a strategy introduced in \cite{Virally:2025nnl} which encodes the Mellin space analogue of $\Delta_\varepsilon$. The authors of \cite{Chester:2025jxg} derived reduced exchange diagrams (which are in one-to-one correspondence with conformal blocks in position space) for almost all supermultiplets appearing in $\langle\phi_{2}\phi_{2}\phi_{k}\phi_{k}\rangle$ and $\langle\phi_{2}\phi_{k}\phi_{2}\phi_{k}\rangle$. The exceptions were in 3d where reduced exchange diagrams for $\DD[2]$ in $\langle\phi_{2}\phi_{2}\phi_{k}\phi_{k}\rangle$ and $\DD[k]$ in $\langle\phi_{2}\phi_{k}\phi_{2}\phi_{k}\rangle$ were not found. The authors attributed this to their definition of the associated superblocks in position space as negative spin limits of certain long blocks, following \cite{Chang:2017xmr}. Our organization in Table \ref{reducedblocktable} encodes these multiplets (named $\DD[p+2]$ in our conventions) using the single-variable reduced blocks which circumvent negative spin limits. It would be nice to understand this reformulation in Mellin space.

Finally, we observe that the reduced correlator decomposition in \eqref{eq:DGSdecomphalfsusy} suffers similar shortcomings to the maximally supersymmetric result in \cite{Woolley:2026cii} when one wishes to derive reduced blocks at higher extremality $\EE$. In particular, the way in which the $\EE-2$ additional reduced correlator functions $b^{\{k_i\}}_I$ appear in the channel equations shows that they are no longer clearly associated with the new multiplets appearing at higher extremality. For instance, \cite{Baume:2019aid} showed that superconformal blocks decompose into a maximum of five $\mathfrak{su}(2)_R$ representations. Correlators with $\EE>4$ will always involve functions $b^{\{k_i\}}_I$ that appear in more than five $\mathfrak{su}(2)_R$ channels, meaning that they are either not clearly associated with particular supermultiplets or that their reduced blocks are intricately tuned to cancel in the remaining channels. The latter cancellations are not possible given the simple reduced blocks we displayed in Equation \eqref{eq:breducedblock}. We feel that an alternative basis of reduced correlators will be needed for convenient reduced block decompositions at higher extremality. The basis-independent solution to the Ward identity appearing in Equation (5.38) of \cite{Dolan:2004mu} will be a useful starting point in this pursuit.

\section*{Acknowledgments}

We are very grateful to Costis Papageorgakis and Shai Chester for encouraging discussions. This project was funded by a Science and Technology Facilities Council (STFC) studentship. 

\begin{appendix}

\section{Conformal block conventions}
\label{superblockconventions}
In this Appendix, we define our conventions for the $\mathfrak{so}(d+1,1)$ conformal blocks $G^{\Delta_{12},\Delta_{34}}_{\Delta,\ell}$ which appear throughout the main text. Conformal blocks encode the contribution of a conformal primary and all conformal descendants and can be defined as solutions to the $\mathfrak{so}(d+1,1)$ conformal quadratic Casimir equation \cite{Dolan:2003hv}. Suppressing the dependence on $\varepsilon$, the conformal Casimir equation takes the form
\begin{align}
    \hat{\CC}^{\Delta_{12},\Delta_{34}}\;G^{\Delta_{12},\Delta_{34}}_{\Delta,\ell}(z,\bar{z})=c_{\Delta,\ell}\;G^{\Delta_{12},\Delta_{34}}_{\Delta,\ell}(z,\bar{z}),
\label{eq:Casimireq}
\end{align}
in terms of the quadratic conformal Casimir operator
\begin{align}
    \hat{\CC}^{\Delta_{12},\Delta_{34}}=D_z+D_{\bar{z}}+2\varepsilon\frac{z \bar{z}}{z-\bar{z}}\left(\left(1-z\right)\partial_z-\left(1-\bar{z}\right)\partial_{\bar{z}}\right),
\end{align}
where we define\footnote{Note that the operator $D_z$ is not to be confused with the operator $D_\varepsilon$ defined in \eqref{eq:Depsilon}.}
\begin{align}
    D_z=z^2\partial_{z}\left(1-z\right)\partial_z+\frac{\Delta_{12}-\Delta_{34}}{2}z^2\partial_z+\frac{\Delta_{12}\Delta_{34}}{4}z.
\end{align}
The Casimir operator has eigenvalues
\begin{align}
    c_{\Delta,\ell}=\frac{1}{2}\left(\ell\left(\ell+2\varepsilon\right)+\Delta\left(\Delta-2\varepsilon-2\right)\right),
\end{align}
which are independent of $\Delta_{ij}$. Analytic solutions to the conformal Casimir equation can be found in even dimensions, such as for when $d=4,6$ in \cite{Dolan:2000ut,Dolan:2003hv}. When $d=4$, we have that 
\begin{align}
&\;G_{\Delta, \ell}^{\Delta_{12}, \Delta_{34}}(z, \bar{z})= \nonumber \\
&\;\frac{(z \bar{z})^{\frac{\Delta-\ell}{2}}}{z-\bar{z}}\left((-z)^{\ell} z\;{ }_2 F_1\left(\frac{\Delta+\ell-\Delta_{12}}{2}, \frac{\Delta+\ell+\Delta_{34}}{2}, \Delta+\ell; z\right)\right. \nonumber \\
&\;\hspace{2cm}\times\left.{}_2 F_1\left(\frac{\Delta-\ell-2-\Delta_{12}}{2}, \frac{\Delta-\ell-2+\Delta_{34}}{2}, \Delta-\ell-2; \bar{z}\right)-(z \leftrightarrow \bar{z})\right).
\end{align}
In $d=6$, we have that 
\begin{align}
&G_{\Delta, \ell}^{\Delta_{12}, \Delta_{34}}(z, \bar{z})= \nonumber \\
&\mathcal{F}_{0,0}-\frac{(\ell+3)}{\ell+1} \mathcal{F}_{-1,1}+\frac{2(\Delta-4) \Delta_{12} \Delta_{34}(\ell+3)}{(\Delta+\ell)(\Delta+\ell-2)(\Delta-\ell-4)(\Delta-\ell-6)} \mathcal{F}_{0,1} \nonumber \\
&+\frac{(\Delta-4)(\ell+3)\left(\Delta-\Delta_{12}-\ell-4\right)\left(\Delta+\Delta_{12}-\ell-4\right)\left(\Delta+\Delta_{34}-\ell-4\right)\left(\Delta-\Delta_{34}-\ell-4\right)}{16(\Delta-2)(\ell+1)(\Delta-\ell-5)(\Delta-\ell-4)^2(\Delta-\ell-3)} \mathcal{F}_{0,2} \nonumber \\
&-\frac{(\Delta-4)\left(\Delta-\Delta_{12}+\ell\right)\left(\Delta+\Delta_{12}+\ell\right)\left(\Delta+\Delta_{34}+\ell\right)\left(\Delta-\Delta_{34}+\ell\right)}{16(\Delta-2)(\Delta+\ell-1)(\Delta+\ell)^2(\Delta+\ell+1)} \mathcal{F}_{1,1},
\end{align}
\noindent
in terms of 
\begin{align}
&\mathcal{F}_{m,n}(z, \bar{z}) = \nonumber \\
&\frac{(z \bar{z})^{\frac{\Delta-\ell}{2}}}{(z-\bar{z})^3}\left((-z)^{\ell} z^{m+3} \bar{z}^n{ }_2 F_1\left(\frac{\Delta+\ell-\Delta_{12}}{2}+m, \frac{\Delta+\ell+\Delta_{34}}{2}+m, \Delta+\ell+2 m ; z\right)\right. \nonumber \\
&\hspace{1.5cm}\times\left.{}_2 F_1\left(\frac{\Delta-\ell-\Delta_{12}}{2}-3+n, \frac{\Delta-\ell+\Delta_{34}}{2}-3+n, \Delta-\ell-6+2 n ; \bar{z}\right)-(z \leftrightarrow \bar{z})\right).
\end{align}
These conventions were chosen such that for any $\varepsilon$, the leading small $z,\bar{z}$ behavior is
\begin{align}
    G_{\Delta, \ell}^{\Delta_{12}, \Delta_{34}}(z, \bar{z})\sim(-1)^\ell z^{\frac{\Delta-\ell}{2}}\bar{z}^{\frac{\Delta+\ell}{2}}\hspace{1cm}\text{for }z\ll\bar{z}\ll1.
\end{align}
Analytic formulae for conformal blocks in odd dimensions are not known, although in Appendix \ref{jackpolys} we will express conformal blocks in general dimensions $d=2(\varepsilon+1)$ in infinite expansions of two-variable symmetric functions known as Jack polynomials. 

As described in \cite{Woolley:2026cii}, $d=2(\varepsilon+1)$ dimensional conformal blocks with twist $t=\Delta-\ell=2\varepsilon-\Delta_{34}$ will turn out to be related to global $SL(2,\mathbb{R})$ block $g^{\Delta_{12},\Delta_{34}}_{\Delta,\ell}$ which we define as
\begin{align}
    g_{\Delta,\ell}^{\Delta_{12},\Delta_{34}}(z)=(-1)^\ell z^{\frac{\Delta+\ell}{2}}{ }_2 F_1\left(\frac{\Delta+\ell}{2}-\frac{\Delta_{12}}{2},\frac{\Delta+\ell}{2}+\frac{\Delta_{34}}{2},\Delta+\ell,z\right).
\label{eq:sl2Rblock}
\end{align}

\section{Jack Polynomials}
\label{jackpolys}
Jack polynomials $P^{(\varepsilon)}_{a,b}$ form a useful basis for expanding symmetric 2-variable functions, such as conformal blocks and the reduced correlators in the main text. They can be defined in terms of Gegenbauer polynomials as
\begin{align}
    P^{(\varepsilon)}_{a,b}(z,\bar{z})=\frac{(a-b)!}{\left(2\varepsilon\right)_{a-b}}\left(z \bar{z}\right)^{\frac{1}{2}\left(a+b\right)}C^{(\varepsilon)}_{a-b}\left(\frac{z+\bar{z}}{2\left(z \bar{z}\right)^{1/2}}\right),
\end{align}
and satisfy an orthogonality condition given by
\begin{align}
    \int^{1}_{-1}C_m^{(\varepsilon)}(x)C_n^{(\varepsilon)}(x)\left(1-x^2\right)^{\varepsilon-\frac{1}{2}}dx=\delta_{m,n}\frac{2^{1-2\varepsilon}\pi\Gamma\left[m+2\varepsilon\right]}{m!(m+\varepsilon)\Gamma\left[\varepsilon\right]^2}.
\end{align}
For $d=4,6$, Jack polynomials can be written explicitly as
\begin{align}
    P^{\left(1\right)}_{a,b}(z,\bar{z})=&\; \frac{1}{\left(a-b+1\right)\left(z-\bar{z}\right)}\left(z^{a+1}\bar{z}^{b}-z^{b}\bar{z}^{a+1}\right),\\
    P^{(2)}_{a,b}(z,\bar{z})=&\;\frac{6}{\left(a-b+2\right)\left(z-\bar{z}\right)^3}\left(\frac{z^{a+3}\bar{z}^{b}-z^b\bar{z}^{a+3}}{a-b+3}-\frac{z^{a+2}\bar{z}^{b+1}-z^{b+1}\bar{z}^{a+2}}{a-b+1}\right).
\end{align}
As discussed in the main text, conformal blocks in $d=2(1+\varepsilon)$ dimensions can be expressed in the expansion
\begin{align}
    G^{\Delta_{12},\Delta_{34}}_{\Delta,\ell}(z,\bar{z})=(-1)^{\ell}\sum_{m=0}^\infty\sum_{n=0}^\infty \;r_{m,n;\Delta,\ell}^{\Delta_{12},\Delta_{34}}\;P^{(\varepsilon)}_{\frac{\Delta+\ell}{2}+m,\frac{\Delta-\ell}{2}+n}(z,\bar{z}).
\label{eq:Gjackpoly}
\end{align}
The action of the conformal quadratic Casimir operator $\hat{\CC}^{\Delta_{12},\Delta_{34}}$ on Jack polynomials
\begin{align}
    \hat{\CC}^{\Delta_{12},\Delta_{34}}\;P^{(\varepsilon)}_{a,b}=&\;\left(a(a-1)+b(b-1-2\varepsilon)\right)P^{(\varepsilon)}_{a,b}\nonumber \\
    -&\;\frac{a-b+2\varepsilon}{a-b+\varepsilon}\left(a-\frac{\Delta_{12}}{2}\right)\left(a+\frac{\Delta_{34}}{2}\right)P^{(\varepsilon)}_{a+1,b}\nonumber \\
    -&\;\frac{a-b}{a-b+\varepsilon}\left(b-\frac{\Delta_{12}}{2}-\varepsilon\right)\left(b+\frac{\Delta_{34}}{2}-\varepsilon\right)P^{(\varepsilon)}_{a,b+1}, 
\end{align}
\noindent
implies that the expansion coefficients $r_{m,n;\Delta,\ell}^{\Delta_{12},\Delta_{34}}$ satisfy
\begin{align}
    &\left(m\left(\Delta+\ell+m-1\right)+n\left(\Delta-\ell+n-1-2\varepsilon\right)\right)r^{\Delta_{12},\Delta_{34}}_{m,n;\Delta,\ell}=\nonumber \\
    &\;\;\;\;\;\;\;\;\frac{\ell+m-n-1+2\varepsilon}{\ell+m-n-1+\varepsilon}\left(\frac{\Delta+\ell-\Delta_{12}}{2}+m-1\right)\left(\frac{\Delta+\ell+\Delta_{34}}{2}+m-1\right)r^{\Delta_{12},\Delta_{34}}_{m-1,n;\Delta,\ell}\nonumber \\
    &\;\;\;\;+\frac{\ell+m-n+1}{\ell+m-n+1+\varepsilon}\left(\frac{\Delta-\ell-\Delta_{12}}{2}+n-1-\varepsilon\right)\left(\frac{\Delta-\ell+\Delta_{34}}{2}+n-1-\varepsilon\right)r^{\Delta_{12},\Delta_{34}}_{m,n-1;\Delta,\ell}.
\end{align}
This recursion relation was solved in \cite{Dolan:2003hv} so that the coefficients take the form
\begin{align}
    r^{\Delta_{12},\Delta_{34}}_{m,n;\Delta,\ell}=&\;\left(\frac{\Delta+\ell-\Delta_{12}}{2} \right)_m \left(\frac{\Delta+\ell +\Delta_{34}}{2}\right)_m \left(\frac{\Delta-\ell -\Delta_{12}-2 \varepsilon}{2}\right)_n \left(\frac{\Delta-\ell +\Delta_{34}-2 \varepsilon}{2}\right)_n\nonumber \\
    \times&\;\frac{(2 \varepsilon )_\ell}{(\varepsilon )_\ell}\frac{(\varepsilon +\ell+m-n) }{(\varepsilon +\ell+m)}
    \frac{ (\varepsilon +1-\Delta )_\ell (\ell+m)! (2 \varepsilon )_{\ell+m-n}}{ m! n! (2 \varepsilon +1-\Delta )_\ell  (\ell+m-n)! (\Delta+\ell )_m (\varepsilon )_{\ell+m} (\Delta-\ell -\varepsilon )_n}\nonumber \\
    \times&\;\, _4F_3\left(
    \begin{matrix}
        \varepsilon, -\ell-m+n,-\ell, \Delta -1\\
        2 \varepsilon, -\ell-m, -\ell+n+\Delta -\varepsilon
    \end{matrix}
;1\right).
\label{eq:rcoeff}
\end{align}
The derivation of reduced superblocks in the main text relied on the following relation that these expansion coefficients obey, namely 
\begin{align}
    &\frac{r_{m,n;\Delta,\ell}^{\Delta_{12},\Delta_{34}}}{\left(\frac{\Delta+\ell}{2}+m+\frac{\Delta_{34}}{2}-\varepsilon+1\right)_{\varepsilon-1}\left(\frac{\Delta-\ell}{2}+n+\frac{\Delta_{34}}{2}-2\varepsilon+1\right)_{\varepsilon-1}}=\nonumber \\
    &\hspace{3cm}\frac{\Gamma \left(\frac{1}{2} \left( \Delta-\ell -4 \varepsilon +\Delta _{34}+2\right)\right) \Gamma \left(\frac{1}{2} \left(\Delta+\ell -2 \varepsilon +\Delta _{34}+2\right)\right)}{\Gamma \left(\frac{1}{2} \left(\Delta+\ell +\Delta _{34}\right)\right) \Gamma \left(\frac{1}{2} \left(\Delta-\ell -2 \varepsilon +\Delta _{34}\right)\right)}\;r_{m,n;\Delta,\ell}^{\Delta_{12},\Delta_{34}-2\left(\varepsilon-1\right)}.
\end{align}
Finally, we note that Jack polynomials
are eigenfunctions of the operator $\Delta_\varepsilon$, satisfying 
\begin{align}
    \Delta_f P^{(\varepsilon)}_{a,b}=(a+\varepsilon+1)_{f-1}(b+1)_{f-1}P^{(\varepsilon)}_{a,b}.
\label{eq:eigenjackpoly}
\end{align}

\end{appendix}

%% Bibliography

%%\bibliography{6dWalgebra}

\printbibliography

@article{Dolan:2004mu,
    author = "Dolan, Francis A. and Gallot, Laurent and Sokatchev, Emery",
    title = "{On four-point functions of 1/2-BPS operators in general dimensions}",
    eprint = "hep-th/0405180",
    archivePrefix = "arXiv",
    reportNumber = "DAMTP-04-45, LAPTH-1047-04",
    doi = "10.1088/1126-6708/2004/09/056",
    journal = "JHEP",
    volume = "09",
    pages = "056",
    year = "2004"
}

@article{Dolan:2000ut,
    author = "Dolan, F. A. and Osborn, H.",
    title = "{Conformal four point functions and the operator product expansion}",
    eprint = "hep-th/0011040",
    archivePrefix = "arXiv",
    reportNumber = "DAMTP-2000-125",
    doi = "10.1016/S0550-3213(01)00013-X",
    journal = "Nucl. Phys. B",
    volume = "599",
    pages = "459--496",
    year = "2001"
}

@article{Dolan:2003hv,
    author = "Dolan, F. A. and Osborn, H.",
    title = "{Conformal partial waves and the operator product expansion}",
    eprint = "hep-th/0309180",
    archivePrefix = "arXiv",
    reportNumber = "DAMTP-03-91",
    doi = "10.1016/j.nuclphysb.2003.11.016",
    journal = "Nucl. Phys. B",
    volume = "678",
    pages = "491--507",
    year = "2004"
}

@article{Dolan:2001tt,
    author = "Dolan, F. A. and Osborn, H.",
    title = "{Superconformal symmetry, correlation functions and the operator product expansion}",
    eprint = "hep-th/0112251",
    archivePrefix = "arXiv",
    reportNumber = "DAMTP-01-82",
    doi = "10.1016/S0550-3213(02)00096-2",
    journal = "Nucl. Phys. B",
    volume = "629",
    pages = "3--73",
    year = "2002"
}

@article{Beem:2013sza,
    author = "Beem, Christopher and Lemos, Madalena and Liendo, Pedro and Peelaers, Wolfger and Rastelli, Leonardo and van Rees, Balt C.",
    title = "{Infinite Chiral Symmetry in Four Dimensions}",
    eprint = "1312.5344",
    archivePrefix = "arXiv",
    primaryClass = "hep-th",
    reportNumber = "YITP-SB-13-45, CERN-PH-TH-2013-311, HU-EP-13-78",
    doi = "10.1007/s00220-014-2272-x",
    journal = "Commun. Math. Phys.",
    volume = "336",
    number = "3",
    pages = "1359--1433",
    year = "2015"
}

@article{Ferrara:2001uj,
    author = "Ferrara, Sergio and Sokatchev, Emery",
    title = "{Universal properties of superconformal OPEs for 1/2 BPS operators in 3 \ensuremath{<}= D \ensuremath{<}= 6}",
    eprint = "hep-th/0110174",
    archivePrefix = "arXiv",
    reportNumber = "CERN-TH-2001-285, LAPTH-872-01",
    doi = "10.1088/1367-2630/4/1/302",
    journal = "New J. Phys.",
    volume = "4",
    pages = "2",
    year = "2002"
}

@article{Dolan:2011dv,
    author = "Dolan, F. A. and Osborn, H.",
    title = "{Conformal Partial Waves: Further Mathematical Results}",
    eprint = "1108.6194",
    archivePrefix = "arXiv",
    primaryClass = "hep-th",
    reportNumber = "DAMTP-11-64, CCTP-2011-32",
    month = "8",
    year = "2011"
}

@article{Chester:2023qwo,
    author = "Chester, Shai M. and Pufu, Silviu S. and Wang, Yifan and Yin, Xi",
    title = "{Bootstrapping M-theory orbifolds}",
    eprint = "2312.13112",
    archivePrefix = "arXiv",
    primaryClass = "hep-th",
    doi = "10.1007/JHEP06(2024)001",
    journal = "JHEP",
    volume = "06",
    pages = "001",
    year = "2024"
}

@article{Chester:2025wti,
    author = "Chester, Shai M. and Mouland, Rishi and van Muiden, Jesse",
    title = "{Extremal couplings, graviton exchange, and gluon scattering in AdS}",
    eprint = "2505.23948",
    archivePrefix = "arXiv",
    primaryClass = "hep-th",
    doi = "10.1007/JHEP10(2025)027",
    journal = "JHEP",
    volume = "10",
    pages = "027",
    year = "2025"
}

@article{Baume:2019aid,
    author = "Baume, Florent and Fuchs, Michael and Lawrie, Craig",
    title = "{Superconformal Blocks for Mixed 1/2-BPS Correlators with $SU(2)$ R-symmetry}",
    eprint = "1908.02768",
    archivePrefix = "arXiv",
    primaryClass = "hep-th",
    reportNumber = "IFT-UAM/CSIC-19-112",
    doi = "10.1007/JHEP11(2019)164",
    journal = "JHEP",
    volume = "11",
    pages = "164",
    year = "2019"
}

@article{Bobev:2017jhk,
    author = "Bobev, Nikolay and Lauria, Edoardo and Mazac, Dalimil",
    title = "{Superconformal Blocks for SCFTs with Eight Supercharges}",
    eprint = "1705.08594",
    archivePrefix = "arXiv",
    primaryClass = "hep-th",
    doi = "10.1007/JHEP07(2017)061",
    journal = "JHEP",
    volume = "07",
    pages = "061",
    year = "2017"
}

@article{Chang:2017xmr,
    author = "Chang, Chi-Ming and Lin, Ying-Hsuan",
    title = "{Carving Out the End of the World or (Superconformal Bootstrap in Six Dimensions)}",
    eprint = "1705.05392",
    archivePrefix = "arXiv",
    primaryClass = "hep-th",
    reportNumber = "CALT-TH-2017-015",
    doi = "10.1007/JHEP08(2017)128",
    journal = "JHEP",
    volume = "08",
    pages = "128",
    year = "2017"
}

@article{Aprile:2021pwd,
    author = "Aprile, Francesco and Heslop, Paul",
    title = "{Superconformal Blocks in Diverse Dimensions and BC Symmetric Functions}",
    eprint = "2112.12169",
    archivePrefix = "arXiv",
    primaryClass = "hep-th",
    doi = "10.1007/s00220-023-04740-7",
    journal = "Commun. Math. Phys.",
    volume = "402",
    number = "2",
    pages = "995--1101",
    year = "2023"
}

@article{Beem:2014zpa,
    author = "Beem, Christopher and Lemos, Madalena and Liendo, Pedro and Rastelli, Leonardo and van Rees, Balt C.",
    title = "{The $ \mathcal{N}=2 $ superconformal bootstrap}",
    eprint = "1412.7541",
    archivePrefix = "arXiv",
    primaryClass = "hep-th",
    reportNumber = "HU-EP-14-61, YITP-SB-14-54, CERN-PH-TH-2014-269, HU-EP-14/61",
    doi = "10.1007/JHEP03(2016)183",
    journal = "JHEP",
    volume = "03",
    pages = "183",
    year = "2016"
}

@article{Chester:2025jxg,
    author = "Chester, Shai M. and Mouland, Rishi and van Muiden, Jesse and Virally, Cl{\'e}ment",
    title = "{Extremal couplings and gluon scattering in M-theory}",
    eprint = "2512.04057",
    archivePrefix = "arXiv",
    primaryClass = "hep-th",
    doi = "10.1007/JHEP06(2026)090",
    journal = "JHEP",
    volume = "06",
    pages = "090",
    year = "2026"
}

@article{Buric:2019rms,
    author = "Buric, Ilija and Schomerus, Volker and Sobko, Evgeny",
    title = "{Superconformal Blocks: General Theory}",
    eprint = "1904.04852",
    archivePrefix = "arXiv",
    primaryClass = "hep-th",
    reportNumber = "DESY 19-057, DESY-19-057, NORDITA 2019-032",
    doi = "10.1007/JHEP01(2020)159",
    journal = "JHEP",
    volume = "01",
    pages = "159",
    year = "2020"
}

@article{Jack:1970,
    author = "Jack, Henry",
    title = "{A class of symmetric polynomials with a parameter}",
    journal = "Proc. Roy. Soc. Edinburgh",
    volume = "69",
    pages = "1",
    year = "1970"
}

@article{Chester:2014mea,
    author = "Chester, Shai M. and Lee, Jaehoon and Pufu, Silviu S. and Yacoby, Ran",
    title = "{Exact Correlators of BPS Operators from the 3d Superconformal Bootstrap}",
    eprint = "1412.0334",
    archivePrefix = "arXiv",
    primaryClass = "hep-th",
    reportNumber = "PUPT-2476, MIT-CTP-4614",
    doi = "10.1007/JHEP03(2015)130",
    journal = "JHEP",
    volume = "03",
    pages = "130",
    year = "2015"
}

@article{Virally:2025nnl,
    author = "Virally, Cl{\'e}ment",
    title = "{Solving superconformal Ward identities in Mellin space}",
    eprint = "2503.09703",
    archivePrefix = "arXiv",
    primaryClass = "hep-th",
    doi = "10.1007/JHEP06(2025)236",
    journal = "JHEP",
    volume = "06",
    pages = "236",
    year = "2025"
}

@article{Woolley:2026cii,
    author = "Woolley, Mitchell",
    title = "{Reduced superblocks at next-to-next-to-extremality for all maximally supersymmetric CFTs}",
    eprint = "2601.15407",
    archivePrefix = "arXiv",
    primaryClass = "hep-th",
    reportNumber = "QMUL-PH-26-02",
    month = "1",
    year = "2026"
}

@article{Alday:2021odx,
    author = "Alday, Luis F. and Behan, Connor and Ferrero, Pietro and Zhou, Xinan",
    title = "{Gluon Scattering in AdS from CFT}",
    eprint = "2103.15830",
    archivePrefix = "arXiv",
    primaryClass = "hep-th",
    doi = "10.1007/JHEP06(2021)020",
    journal = "JHEP",
    volume = "06",
    pages = "020",
    year = "2021"
}

@article{Chang:2019dzt,
    author = "Chang, Chi-Ming and Fluder, Martin and Lin, Ying-Hsuan and Shao, Shu-Heng and Wang, Yifan",
    title = "{3d N=4 Bootstrap and Mirror Symmetry}",
    eprint = "1910.03600",
    archivePrefix = "arXiv",
    primaryClass = "hep-th",
    reportNumber = "IPMU19-0116, CALT-TH 2019-024, PUPT-2596",
    doi = "10.21468/SciPostPhys.10.4.097",
    journal = "SciPost Phys.",
    volume = "10",
    number = "4",
    pages = "097",
    year = "2021"
}

@article{Chang:2017cdx,
    author = "Chang, Chi-Ming and Fluder, Martin and Lin, Ying-Hsuan and Wang, Yifan",
    title = "{Spheres, Charges, Instantons, and Bootstrap: A Five-Dimensional Odyssey}",
    eprint = "1710.08418",
    archivePrefix = "arXiv",
    primaryClass = "hep-th",
    reportNumber = "CALT-TH-2017-030, PUPT-2539",
    doi = "10.1007/JHEP03(2018)123",
    journal = "JHEP",
    volume = "03",
    pages = "123",
    year = "2018"
}

@article{Behan:2024vwg,
    author = "Behan, Connor and Chester, Shai M. and Ferrero, Pietro",
    title = "{Towards bootstrapping F-theory}",
    eprint = "2403.17049",
    archivePrefix = "arXiv",
    primaryClass = "hep-th",
    doi = "10.1007/JHEP10(2024)161",
    journal = "JHEP",
    volume = "10",
    pages = "161",
    year = "2024"
}

\end{document}